\documentclass[usenatbib]{mn2e}

\usepackage{graphicx}
\usepackage{amssymb}
\usepackage{epsfig}

\title[SXP7.92]{An X-ray and optical study of the new SMC X-ray binary pulsar system SXP7.92 and its probable optical counterpart, AzV285.}

\author[M.J. Coe et al.]{M. J.~Coe$^{1}$,
 M. ~Schurch$^{1}$, V.A. ~McBride$^{1}$, R.H.D. ~Corbet$^{2}$, L.J. Townsend $^{1}$, \and A. Udalski$^{3}$, J.L. Galache$^{4}$\\
$^{1}$ School of Physics and Astronomy, University of Southampton, SO17
1BJ, UK\\
$^{2}$ University of Maryland, Baltimore County, Mail Code 662, NASA Goddard Space Flight Center, Greenbelt, MD 20771, USA \\
$^{3}$ Warsaw University Observatory, Aleje Ujazdowskie 4, 00-478 Warsaw, Poland \\
$^{4}$ Harvard-Smithsonian Center for Astrophysics, 60 Garden Street,
Cambridge, MA 02138, USA. \\}

\begin{document}

\date{7 Jan 2009}

\pagerange{\pageref{firstpage}--\pageref{lastpage}} \pubyear{2002}

\maketitle

\label{firstpage}

\begin{abstract}

Optical and X-ray observations are presented here of a newly reported X-ray transient system in the Small Magellanic Cloud - SXP7.92. A detailed analysis of the X-ray data reveal a coherent period of 7.9s. A search through earlier X-ray observations of the SMC reveal a previously unknown earlier detection of this system. Follow-up X-ray observations identified a new transient source within the error circle of the previous observations. An optical counterpart, AzV285, is proposed which reveals clear evidence for a 36.8d binary period.

\end{abstract}


\section{Introduction and background}

The Be/X-ray systems represent the largest sub-class of all High Mass X-ray Binaries (HMXB).  A survey of the literature reveals that of the $\sim$240 HMXBs known in our galaxy and the Magellanic Clouds (Liu et al., 2005, 2006), $\ge$50\%
fall within this class of binary.  The orbit of the Be star
and the compact object, presumably a neutron star, is generally wide
and eccentric.  X-ray outbursts are normally associated with the
passage of the neutron star close to the circumstellar disk (Okazaki
\& Negueruela 2001). A review of such systems may
be found in Negueruela (1998) and Coe et al. (2000).

The source that is the subject of this paper was first identified by Corbet et al. (2008) in the Small Magellanic Cloud (SMC) using the Rossi X-ray Timing Explorer (RXTE). It has a pulse period of 7.92s so it is designated here as SXP7.92 following the naming convention of Coe et al. (2005). Follow up observations by the Swift X-ray observatory identified a new X-ray source within the RXTE field of view, but the weakness of the X-ray signal precluded searching for the 7.92s pulsations. In this paper over 10 years of optical photometry data from the Optical Gravitational Lensing Experiment (OGLE) of the probable optical counterpart at the Swift source position are presented. On occasion these optical data reveal a clear 36.8d modulation which is interpreted here as the binary period of the system.

\section{X-ray data}

\subsection{RXTE}

The SMC has been the subject of extensive monitoring using RXTE Proportional Counter Array (PCA) over the last 10 years (Laycock et al., 2005, Galache et al., 2008), and
SXP7.92 was discovered in an observation taken 13 June 2008. This observation was centred on RA 14.3$^{\circ}$ Dec -72.3$^{\circ}$ with a Full Width Half Maximum (FWHM) sensitivity response field of view of $1\,^{\circ}$. Based upon many recent similar detections of known Be/X-ray sources, it is very probable that the source of the X-ray signal lies within this FWHM zone, though the larger area with a $2\,^{\circ}$ radius down to zero response cannot be excluded.

\begin{figure*}
\begin{center}
\includegraphics[width=100mm,angle=90]{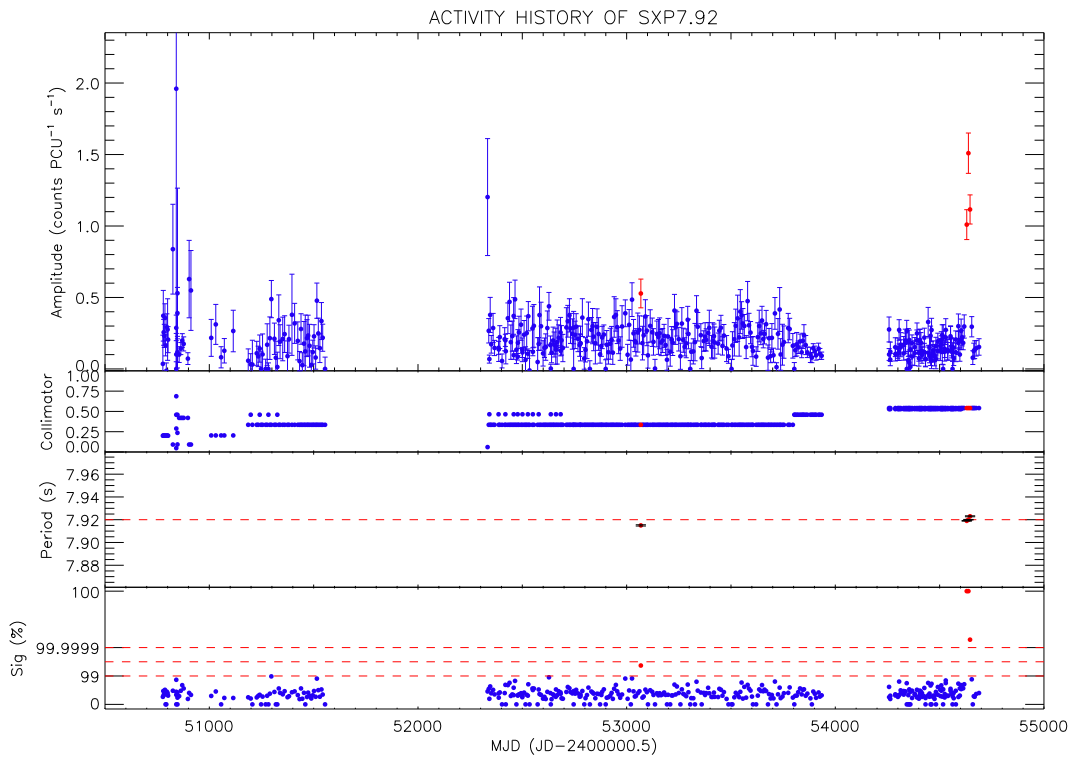}
\caption{RXTE light curve of SXP7.92. The top panel shows the pulse amplitude; the next panel down the RXTE collimator response to SXP7.92; the following panel is the detected pulse period; the lowest panel shows the significance of the pulse period detection. The most significant detections are shown in red.}
\label{fig:xte}
\end{center}
\end{figure*}

Since a data base of $\ge$10 years of observations of the SMC exist it was therefore possible to use these data to search for evidence of any other X-ray detections of SXP7.92 during this period - see Figure~\ref{fig:xte}.  Because the PCA is a collimated instrument, interpreting the strength and significance of the signal depends upon the position of the source within the field of view. So the target was assumed to be located at the position of the possible counterpart determined from the SWIFT observation (see next section). Apart from the set of recent detections which led to the identification of this new pulsar, there is only one previous convincing detection of SXP7.92 on 3 March 2004 (MJD53068).

The search was based around a pulse period of 7.92s with a $\pm$0.05s search window. There are two other pulsars within the same RXTE field of view with similar periods: the proposed Anomalous X-ray Pulsar (AXP) CXOU J010043.1-721134 (McGarry et al., 2005) and SMC X-3. For most of the observations reported here SXP7.92 had a collimator response of 0.55, the AXP source had a collimator response of 0.70 and SMC X-3 that of 0.55.

McGarry et al. (2005) quote the period of the AXP as 8.020s in Jan 2004 from Chandra observations. Our earliest measurement was taken in March 2004 and is 7.915s. McGarry et al. also quote a period change of $1.88.10^{-11}$ s/s between 2001 and 2004 so there would have been a negligible period change in the AXP in the 2 months between the Chandra observations and our RXTE observation. Therefore none of our measurements, particularly that in March 2004 can be of the AXP. For SMC X-3 the well established period is 7.78s and this is outside the search range used here. The search range of $\pm$0.05s was chosen to avoid picking up either of these other two objects.

\begin{figure}
\begin{center}
\includegraphics[width=60mm,angle=-0]{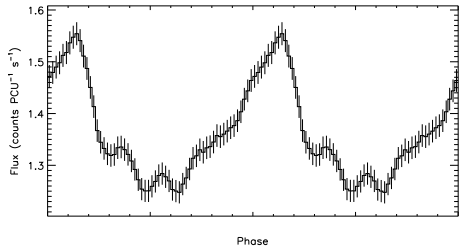}
\caption{Folded pulse profile for observation dated MJD53068.035 (March 2004).}
\label{fig:xte1}
\end{center}
\end{figure}

A list of all the possible detections of SXP7.92 are presented in Table~\ref{tab:rxte}. From this table it is possible to see that there are four detections at the 99.9\% level or higher - the one in 2004 and three recent detections. Therefore these data were used to produce folded pulse light curves.
The folded pulse profile for the observation dated MJD53068 is shown in Figure~\ref{fig:xte1}.

\begin{figure}
\begin{center}
\includegraphics[width=60mm,angle=-0]{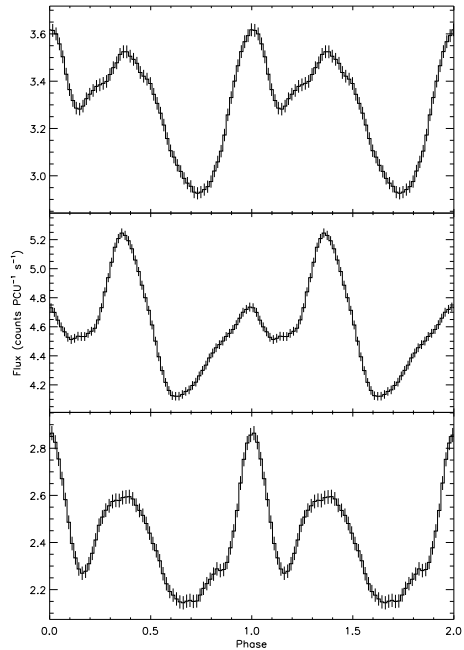}
\caption{Folded RXTE pulse profile for observations dated MJD 54630, 54637 \& 54645, (June 2008) starting with the earliest at the top. The three profiles have been arbitrarily shifted in phase in this figure - see text for discussion.}
\label{fig:xte2}
\end{center}
\end{figure}

All three pulse profiles for observations 93037-01-49 to 51 dated MJD54630-54645 are shown together in  Figure~\ref{fig:xte2} for direct comparison. It must be noted that the pulse phases shown for each of the profiles is arbitrary since it is not possible to phase-link observations so separated in time. So the three profiles shown in Figure~\ref{fig:xte2} have simply been aligned based upon the asymmetric double peak structure. Other configurations cannot be excluded.

The most notable feature of the pulse profiles is the change from an essentially single-peaked outburst in 2004 to the double-peaked structure seen in 2008. This change is probably related to the X-ray luminosity and hence the accretion modes on to the neutron star. Figure~\ref{fig:xte1} shows emission dominated by a thin beam, while Figure~\ref{fig:xte2} shows emission dominated by fan beams. If fan beams arise from an accretion mound with a significant height, we would expect them to appear at higher accretion rates (thus higher $L_{x}$).
And vice versa, at low accretion rates (low $L_{x}$) the emission region will be a relatively flat "spot", that would produce a pencil beam. This is in agreement with the observations presented in this paper if it is assumed that the pulse fraction is approximately constant and the pulsed count rates in Table~\ref{tab:rxte} are representative of the unpulsed flux.

\begin{table}
\begin{center}
\caption{RXTE detections of SXP7.92. The last column ("collimator response") represents the fractional sensitivity of RXTE to the proposed counterpart, AzV285; the centre of the RXTE field of view has response = 1.0.}
\label{tab:rxte}
\begin{tabular}{ccccc} \hline
MJD   &     Amp. &  Signif.    &   Period       &   Coll. \\
&(cts/cpu/s)&\%&s&Resp. \\
53068.0352 &   0.53$\pm$0.01 &  99.9 &    7.915$\pm$0.005 &   0.33 \\
54617.8438 &   0.29$\pm$0.068 & 86.4 &    7.931$\pm$0.003 &   0.54 \\
54630.0000 &   1.01$\pm$0.10 & 100.0 &    7.919$\pm$0.003 &   0.54 \\
54637.9062 &   1.51$\pm$0.14 & 100.0 &    7.920$\pm$0.002 &   0.54 \\
54645.7656 &   1.12$\pm$0.10 &  99.9 &    7.923$\pm$0.004 &   0.54 \\
54653.9492 &   0.30$\pm$0.08 &  97.1 &    7.928$\pm$0.003 &   0.54 \\

\hline
\end{tabular}
\end{center}
\end{table}

\subsection{Swift}

The RXTE PCA instrument has a 2 degree FWZI field of view and hence is not very effective at localising the position of SXP7.92. Consequently the Swift observatory was used to attempt to locate the new X-ray pulsar by searching the region within the RXTE FWHM zone. Though the region outside the RXTE FWHM zone cannot be excluded, the sensitivity of RXTE is such that it would have to be an exceptionally bright X-ray source to have produced the level of signal observed if it fell outside of the FWHM region.  On 16 July 2008 nine observations of approximate duration 1\,ks each, covering the RXTE FWHM zone were made with the X-ray Telescope (XRT) on board Swift in Photon Counting mode.



A single source was discovered at position 01:01:55.3, -72:32:42.8 (J2000.0) with a $\sim4"$ (90\%) error circle as determined through {\it xrtcentroid}. During the 1\,ks exposure time $\sim$200 source counts were detected.

Data in XRT photon counting mode are binned at a time resolution of 2.51 \,s.  A Lomb-Scargle (Lomb 1976, Scargle 1982) analysis in the range 5--1000\,s did not reveal any significant periodicities.  An epoch-folding search (Leahy et al. 1983)  around 7.9\,s (the period measured from the RXTE data) also did not show features other than a strong signal at 3 times the data sampling frequency.

\begin{figure}
\begin{center}
\includegraphics[width=80mm,angle=-0]{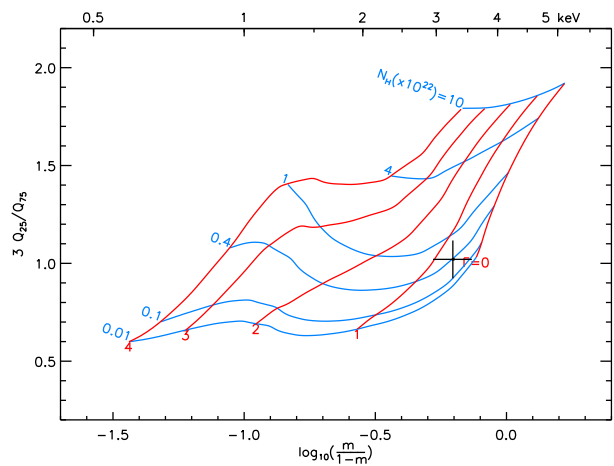}
\caption{Quantile-based colour-colour diagram based on the
median (m) and the ratio of the two quartiles showing the grid
patterns for a power-law model in a 0.5 –- 8.0 keV range ideal detector
for Chandra. The power-law grid patterns are
for $\Gamma$ = 4, 3, 2, 1 and 0 and $N_{H} = 10^{20}, 10^{21}, 4 \times 10^{21}, 10^{22},
4 \times 10^{22}$ and $10^{23} cm^{-2}$. The position of the Swift source on this grid is indicated by the cross symbol.}
\label{fig:nsrc1}
\end{center}
\end{figure}

Using the quantile analysis method of Hong, Schlegel \& Grindlay (2004), we divided the photon distribution of the above source into quartiles.  The energy boundaries of the quartiles, together with the median energy of the distribution were then used to probe the X-ray colours of the source.  Figure~\ref{fig:nsrc1} shows the quantile `colours' with a grid of power law index and absorption overplotted.
From this diagram we can interpret the X-ray spectrum of the source to
have a power law index of $\Gamma\sim0.5$ with an absorption of $N_{\rm H}\sim4\times10^{21}$\,cm$^{-2}$.

\section{Optical and IR data}

\begin{figure}
\begin{center}
\includegraphics[width=70mm,angle=-90]{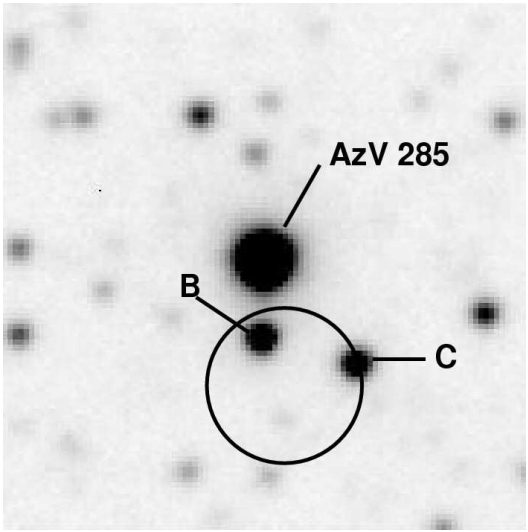}
\caption{OGLE I-band image of the $\sim$25" region near the Swift X-ray source location. The circle indicates the Swift 90\% confidence region.}
\label{fig:fc}
\end{center}
\end{figure}

At first glance the Swift source does not seem to be obviously coincident with any appropriately bright (V$\sim$14-16 - see Coe et al., 2005) objects in the SMC, with the nearest candidate counterparts being V$\sim$17th mag - see Figure~\ref{fig:fc}.  The nearest ``bright'' source is the early-type star AzV285 (Azzopardi, Vigneau \& Macquet 1975) which has B=13.9 mag, K=13.6 mag and is 6.6" away from the centre of the Swift X-ray position.

In total three objects were examined as possible counterparts to the X-ray source. These three objects are shown in Figure~\ref{fig:fc} and their parameters listed in Table~\ref{tab:obj}.

\begin{table}
\begin{center}
\caption{Optical candidates to the X-ray source SXP7.92. Colours are from Zaritsky et al.(2002)}
\label{tab:obj}
\begin{tabular}{ccccl} \hline
Name & RA & Dec & Colour & OGLE III\\
 & &  & U-B & identification\\
&&&&\\
AzV 285 &01 01 56.0&-72 32 36.0&-1.21&SMC110.5 6\\
Object B&01 01 55.8&-72 32 40.2&+0.33&SMC110.5 100\\
Object C&01 01 54.8&-72 32 41.4&+-1.10&SMC110.5 95\\

\hline
\end{tabular}
\end{center}
\end{table}
The two object marked B \& C in Figure~\ref{fig:fc} first need considering as possible counterparts since they are closer to the centre of the Swift position. The colours of all three objects are given in Zaritsky et al.(2002). These are U-B=-1.21 for AzV285, U-B=+0.33 for Object B and U-B=-1.10 for Object C. The colours for Object B are very different from that usually observed for the counterpart to a HMXB in the SMC - typically values of -1.1 to -1.3 are expected for an early type B star. So this makes Object B an unlikely candidate. However, the values for AzV285 and Object C are in this range and they are therefore plausible candidates.

It is also noteworthy that there is an IR source at the location of AzV285 that appears in three different publications, see Table~\ref{tab:obs}. In the 2MASS catalogue it is identified as 2MASS J01015579-7232363, whereas the Simultaneous three colour InfraRed Imager for Unbiased Surveys (SIRIUS) catalogue (Kato et al., 2007) identifies it as 01015579-723264. All three data sets presented in the table show no evidence for variability, except perhaps in the K band.
So, from the 2MASS and SIRIUS data the average IR colour may be determined to be J-K=0.13$\pm$0.05, which would be one of the lowest values for any of the previously identified SMC Be/X-ray binaries (Coe et al., 2005).

\begin{table}
\begin{center}
\caption{IR detections of SXP7.92. Refs are: ZSW = Zickgraf, Stahl \& Wolf (1992); 2MASS=Skrutskie et al. (2006); SIRIUS= Kato et al. (2007)}
\label{tab:obs}
\begin{tabular}{ccccc} \hline
Date & J & H & K & Ref\\
&&&&\\
17 Sep 1989&13.72$\pm$0.12&14.01$\pm$0.29&$\ge$14.0&ZSW \\
8 Sep 1998&13.84$\pm0.03$& 13.81$\pm$0.04&13.66$\pm$0.04&2MASS \\
16 Sep 2002&13.83$\pm$0.02&13.80$\pm$0.02&13.75$\pm$0.02&SIRIUS \\

\hline
\end{tabular}
\end{center}
\end{table}

This region has been monitored by Optical Gravitational Lensing Experiment (OGLE) and more than 10 years worth of regular photometric observations exist. Hence OGLE II \& III data were obtained for all three objects marked in Figure~\ref{fig:fc}. These data were then subjected to a Lomb-Scargle search for any possible periodic modulation that could be indicative of a binary period.

\begin{figure}
\begin{center}
\includegraphics[width=60mm,angle=-90]{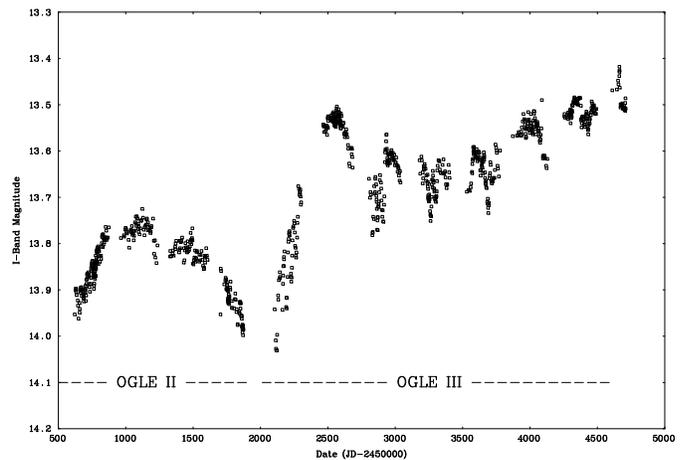}
\caption{I-band photometric data from OGLE II \& III of AzV285.}
\label{fig:ogle}
\end{center}
\end{figure}

The results from OGLE II and OGLE III of AzV285 are shown in Figure~\ref{fig:ogle}. It is immediately apparent that the object is extremely variable in the I band and, at the time of the X-ray outburst of July 2008 it was at its brightest for the last decade. If this object is indeed a Be star, then the variability seen in Figure~\ref{fig:ogle} is attributable to emission from the circumstellar envelope. Hence the current bright state is indicative of the disk being larger than at any time in the last 10 years.

\begin{figure}
\begin{center}
\includegraphics[width=80mm,angle=-0]{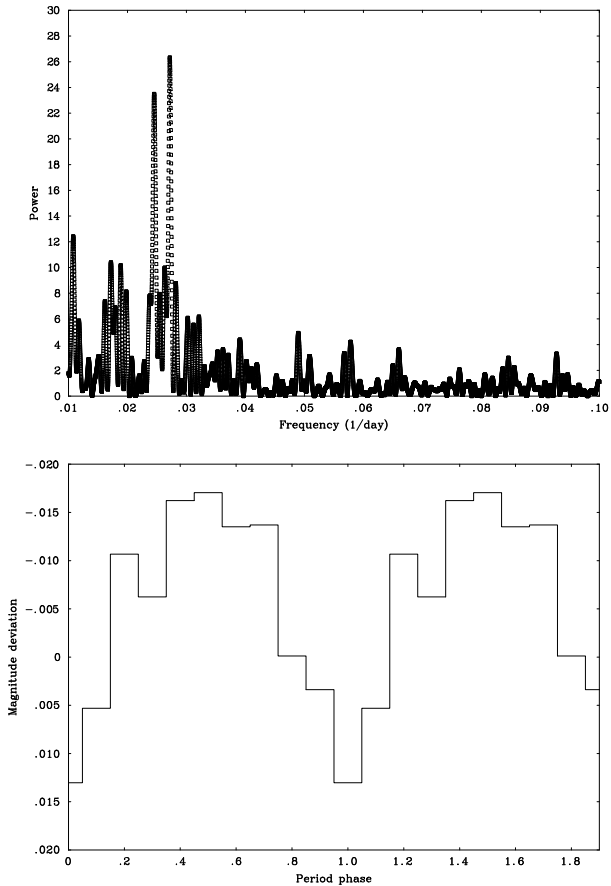}
\caption{Upper panel: Lomb-Scargle power spectrum of detrended OGLE II data of AzV285. The highest peak is at a period of 36.79d. Lower panel: deviations from the detrended OGLE II data folded at the period of 36.79d.}
\label{fig:ogle2_ps}
\end{center}
\end{figure}

The extremely rapid variability on occasions, particularly in the OGLE III data, makes searching for possible binary periods technically difficult. However the earlier, OGLE II data are more amenable to such a search. If this section of data is first de-trended using a simple polynomial fit and then subjected to a Lomb-Scargle frequency search, a strong coherent period is seen at 36.79d - see Figure~\ref{fig:ogle2_ps}. There is second peak at 40.8d which represents the beating between the true period and the annual sampling.

\begin{figure}
\begin{center}
\includegraphics[width=80mm,angle=-0]{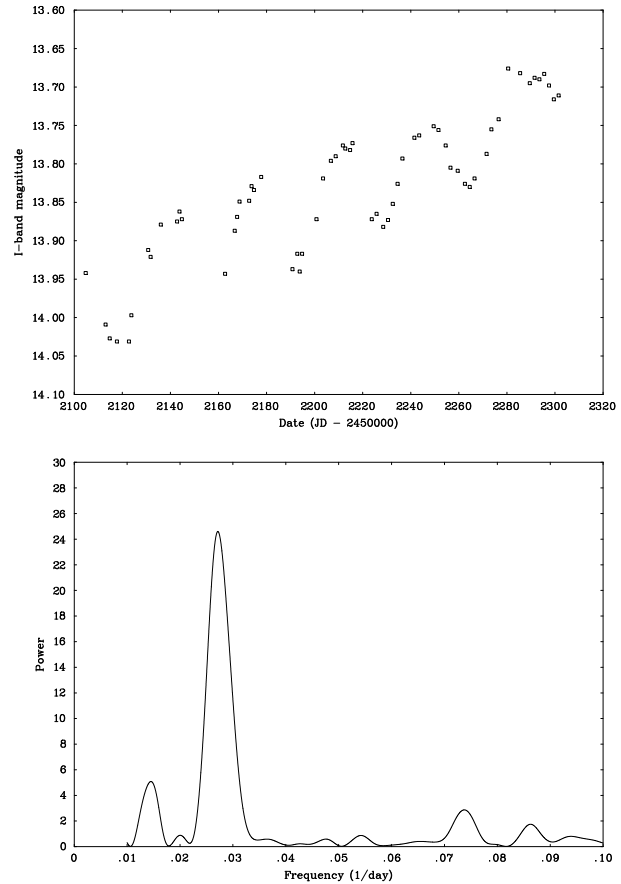}
\caption{Top panel: enlargement of the first year of OGLE III data of AzV285. Lower panel: the Lomb-Scargle power spectrum of this data set. The peak is at a period of 36.83d.}
\label{fig:og4}
\end{center}
\end{figure}

A close look at the first year of OGLE III data also reveals a clear visible modulation even though the the source is brightening very fast - see the top panel of Figure~\ref{fig:og4}. Subjecting this year's worth of data to a Lomb-Scargle analysis strongly confirms the same period seen in the OGLE II data - see lower panel of Figure~\ref{fig:og4}. In addition, the shape of the modulation is very similar to that seen in the folded OGLE II data.

It is worth noting, in passing, that there also exist MACHO data for AzV285 covering the time span JD 2449372 - 2451546. The MACHO source ID is 207.16714.4. These data have a poorer signal to noise ratio than the OGLE III data, but nonetheless reveal a similar long-term brightness change over several years.

A similar detailed timing analysis of the OGLE III data on Object C revealed a strong sinusoidal modulation at a period of 3.3d. The U-B colours of this object make it an unlikely counterpart to the X-ray source (if it is a HMXB). There are also occasional V band data from OGLE III and the mean V-I color of this object is V-I$\sim$0.1 mag. At I$\sim$16.65 this means that the object is on the main sequence of the SMC field stars rather than close to the instability strip and hence a Cepheid variable is ruled out. It is more likely that this is a low amplitude pulsating B-type main sequence star.

Finally, a timing analysis of the OGLE III data for Object B reveals no evidence for any periodic modulation or variability. Though its U-B=-1.10 colours are similar to that of a $\sim$ B0V star, its brightness at V=16.7 is two magnitudes fainter than expected. So it is unlikely to be the counterpart to a Be/X-ray system.

The conclusion, therefore, from the optical data is that AzV285 is a strong contender to be the counterpart to the Swift X-ray source, and hence a new HMXB system.

\section{Discussion}

\subsection{Is AzV285 the optical counterpart to SXP7.92?}

Though the body of evidence presented here is strongly indicative that AzV285 is the optical counterpart to SXP7.92 it is important to point out that the link is not certain. Firstly the pulsar SXP7.92 is only located for sure within the RXTE PCA field of view. Based upon the average observed luminosity of Be/X-ray systems in the SMC whilst undergoing an outburst, it makes it unlikely that the source could be outside of the FWHM field of view, but not impossible.

The Swift search of the RXTE FWHM field revealed just one source, and this source is clearly transient in nature since an earlier XMM observation failed to detect it in June 2007 (JD 2454258). However, the $\sim$200 counts accumulated from the Swift observation were not sufficient to confirm or deny the existence of pulsations at 7.92s. It is, though, clear that the Swift source position is almost certainly compatible with that of the early type star AzV285. So, at the least, this Swift object is very likely a new HMXB system in the SMC - even if it turns out not to be SXP7.92.

\subsection{Spectral class}

It is possible to investigate the probable spectral class of AzV285. However, it is always difficult determining the spectral class of young emission line systems from just photometric magnitudes, specially because the redder values are likely to be affected by the circumstellar disk if the object is indeed a Be star. However, if just the U \& B values are used, then comparison with Kurucz stellar models indicate a spectral class between O9 and B0.

The U-band magnitude of 12.8 is bright compared to a Be star in the SMC.
Assuming a distance modulus to the SMC of 18.9 and a reddening of E(B-V)=0.09 (Schwering \& Israel 1997), then a B0V star would have U = 13.8 a whole magnitude fainter than that reported. The same is true for the observed V magnitude of 13.96 which is a magnitude brighter than that the value of 14.9 expected for a B0V star. Changing the luminosity class to III would produce the extra magnitude needed, as would moving to a much earlier stellar type. But the colours of the earlier type would not match the observed U-B colour. So the best estimate of spectral type based upon the photometry is O9-B0 III. Further refinements will require high quality blue spectra.

Such a classification would put this object at the edge of the spectral distribution of such Be/X-ray binary systems - see McBride et al. (2008) - but not exceptionally so. It would also be, optically, the brightest Be/X-ray system in the SMC.

\subsection{X-ray characteristics}

\begin{figure}
\begin{center}
\includegraphics[width=40mm,angle=-90]{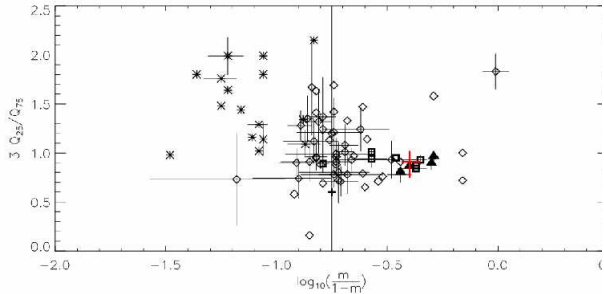}
\caption{Quantile analysis diagram comparing the candidate Swift source (shown in red) to other objects seen in Chandra observations of the SMC (from McGowan et al., 2008). The objects
identified as stars are marked with asterisks, AGNs with diamonds
and HMXBs with filled triangles. The SMC Bar pulsars from Edge
et al. (2004) are also included on the plots, marked with squares. }
\label{fig:nsrc2}
\end{center}
\end{figure}

XMM-Newton observed the field of the Swift X-ray source on 2007 June 6 (Haberl, Eger \& Pietsch 2008).  The 28.9 ks observation showed a number of active X-ray sources in the field, however, no activity was present at the position of the recently detected Swift source.  The transient nature exhibited by the source means we can immediately dismiss the possibility that it may be a background galaxy.  Together with the fact that the majority of SMC X-ray point sources are Be/X-ray binaries, this evidence makes it very likely that the Swift source is a high mass X-ray binary system containing a Be counterpart.

Assuming the Swift source is located in the SMC, the 0.3--10\,keV flux converts into a luminosity of $7.7\times10^{36}$\,erg\,s$^{-1}$, using a distance estimate of 60\,kpc (Harries, Hilditch \& Howarth 2003).  This is within the range of expected values for a Type I outburst of a Be/X-ray binary (Stella et al. 1986).  The spectral parameters, as derived from the quantile analysis in Section 2.2, are consistent with those of other known Be/X-ray binary systems in the SMC (McGowan et al. 2008).

All known Be/X-ray binary systems harbour neutron star counterparts.
Thus we expect to see X-ray pulsations from this object.  The lack of detected pulsations  was explored by performing Monte Carlo simulations of a lightcurve with the same time sampling as that of the XRT data.  A sinusoidal signal of period 7.92\,s and varying pulsed fraction was generated using Poisson statistics, in such a way that means of the simulated lightcurves were consistent with the mean of the real data.
1000 simulations were performed for pulse fractions of 0, 0.1, 0.2 and 0.3, with the 0 pulsed fraction (i.e. Poisson noise only) simulated lightcurves used to define the 90\% significance for the detection of a periodic signal.  With a pulsed fraction of 0.1, 0.2 and 0.3 only 26,
497 and 952 respectively out of 1000 simulated lightcurves showed a significant ($>90\%$) period at 7.92\,s.  These simulations suggest that it would only be likely to detect a periodicity of 7.92\,s in this source if the pulse fraction were around 0.3.  Judging from the PCA pulse profile (Figure~\ref{fig:xte2}), which covers a similar energy range as that of the XRT lightcurve, the pulsed fraction is typically $\sim0.2$.  It may be, therefore, that the combination of the low pulsed fraction and the low signal level prevent any pulsations being detected in the Swift/XRT lightcurve.

\section{Conclusions}

If the source discovered by Swift is SXP7.92, and the Swift source counterpart is indeed AzV285, the orbital and pulse periods place this source well within the distribution of other Be/X-ray binary systems (Corbet et al. 1999).  Future observations which will provide both a better X-ray position, as well as confirmation of a 7.92\,s periodicity, are required to make a conclusive classification of this new X-ray source.

\section{Acknowledgements}

We are grateful to the referee for helpful comments and to the Swift team for scheduling and supporting the follow-up observations. LJT acknowledges the support of a University of Southampton Mayflower Scholarship. The OGLE project is partially supported by the Polish MNiSW grant N20303032/4275.

\bsp

\label{lastpage}

\end{document}